\documentclass[aps,prl,floatfix,twocolumn,nofootinbib,superscriptaddress,preprintnumbers]{revtex4-2}

\usepackage{graphicx}
\usepackage{comment}
\usepackage{dcolumn}
\usepackage{bm}
\usepackage{physics}
\usepackage{amsmath,amssymb,xcolor,float,mathrsfs} 
\usepackage{xparse,etoolbox}
\usepackage{etaremune}
\usepackage{enumitem}
\usepackage[colorlinks,allcolors=blue]{hyperref} 
\usepackage[capitalize]{cleveref}
\usepackage[normalem]{ulem}

\newcommand{\submitletter}{}

\newcommand{\compile}{}

\begin{document}

\ifdefined\compile

\preprint{RESCEU-6/25}

\title{Universal profile for cosmic birefringence tomography with radio galaxies}

\author{Fumihiro Naokawa}
\affiliation{Department of Physics, Graduate School of Science, The University of Tokyo, Bunkyo-ku, Tokyo 113-0033, Japan}
\affiliation{Research Center for the Early Universe, The University of Tokyo, Bunkyo-ku, Tokyo 113-0033, Japan}

\begin{abstract}

We propose a new method to tomographically probe cosmic birefringence using radio galaxies. We show that the redshift evolution of the cosmic birefringence angle induced by a slow-rolling pseudoscalar field, which is a candidate for dynamical dark energy, is independent of the detailed model of the pseudoscalor field. This universal profile evolves predominantly at $z\lesssim10$. In contrast, if the origin is a dark matter-like pseudscalor field, the resulting birefringence angle tends to be negligible in the low-redshift regime. This new insight provides a strong motivation to independently test the cosmic birefringence using polarized astrophysical sources such as radio galaxies. We find that a sample size of $\order{10^5-10^6}$ is required to distinguish the profiles, which is achievable with ongoing and upcoming radio surveys such as ASKAP or SKA. 
\end{abstract}

\maketitle

\fi

\ifdefined \submitletter
  {\it Introduction.---}
\else
  \section{Introduction} \label{sec:intro}
\fi
Cosmic birefringence \cite{Komatsu:2022nvu}, the rotation of the polarization plane of photons during their propagation through the universe, has been investigated as a probe of global parity violation in the universe. This can be induced by a pseudoscalar field $\phi$ such as an ``axion-like" particle, which is a promising candidate for dark energy or dark matter. Recently, Refs.~\cite{Minami:2020odp, Eskilt:2022cff, Eskilt:2022wav, Eskilt:2023ndm, Diego-Palazuelos:2022dsq} have analyzed the Cosmic Microwave Background (CMB) polarization data of Planck and WMAP and found a hint on cosmic birefringence. The tightest constraint on the rotation angle to date is $0.34\pm0.09~$deg and excluding zero rotation at a significance level of 3.6$\sigma$ \cite{Eskilt:2022cff}. The latest data release from the Atacama Cosmology Telescope (ACT) reported the angle as $\beta_0=0.20\pm0.08$ deg \cite{ACT:2025fju} consistent with Planck \& WMAP, though the possibility of systematics still remains as discussed later. 


These hints of cosmic birefringence provide excellent motivation to explore parity-violating physics beyond the standard framework \cite{Fujita:2020ecn,Takahashi:2020tqv,Fung:2021wbz,Nakagawa:2021nme,Jain:2021shf,Choi:2021aze,Obata:2021nql,Nakatsuka:2022epj,Lin:2022niw,Gasparotto:2022uqo,Lee:2022udm,Jain:2022jrp,Murai:2022zur,Gonzalez:2022mcx,Qiu:2023los,Eskilt:2023nxm,Namikawa:2023zux,Gasparotto:2023psh,Ferreira:2023jbu, Yin:2024hba, Tada2024, Lee:2025yvn, Nakagawa:2025ejs}, as it is extremely difficult to explain the reported angle within the Lambda Cold-Dark-Matter($\Lambda$CDM) model or the standard model of particle physics \cite{Nakai:2023zdr}. However, a measurement of $\beta_0$ with the CMB is still challenging. For example, WMAP \& Planck results may still include unknown systematics from foreground emissions. The ACT result might include the systematics from optical systems. Also, the CMB alone can provide limited information on the time evolution of cosmic birefringence, which is critical to determining the origin \cite{Fujita:2020ecn, Nakatsuka:2022epj, Gasparotto:2022uqo}. Therefore, testing the reported signal without the CMB is crucial for further investigation of cosmic birefringence and its origins.

In this Letter, we propose a new method for the tomographic test of cosmic birefringence at the late universe with astrophysical sources such as polarized radio galaxies. This method enables us to test cosmic birefringence induced by the dark energy-like pseudoscalor, which the CMB alone cannot. We show that there is a universal profile of the time evolution of cosmic birefringence by slow-rolling pseudoscalor. We also show that the emerging catalogs of polarized sources can detect cosmic birefringence through the profile in the near future, without relying on the CMB. 

\ifdefined \submitletter
  {\it Cosmic birefringence.---}
\else
  \section{Cosmic birefringence} \label{sec:intro}
\fi
The interaction term between the pseudoscalar field and the photons, known as the Chern-Simons term, $\mathcal{L}\supset-g\phi F_{\mu\nu}\tilde{F}^{\mu\nu}/4$, can produce a rotation of the polarization plane. The total birefringence angle at an arbitrary redshift $z$ is given by
\begin{equation}
    \beta(z) 
    =\frac{1}{2}g\int^{\phi(0)}_{\phi(z)}d\phi 
    =\frac{1}{2}g\left[\phi(0)-\phi(z)\right].
\end{equation}
The dynamics of $\phi$ are governed by the Klein-Gordon equation,
\begin{equation}
\label{24010720}
    \ddot{\phi}+3H(t)\dot{\phi}+V_{,\phi}=0,
\end{equation}
where $H(t)$ is a Hubble parameter, $V$ is the potential of $\phi$, an overdot is time derivative, and the subscript $_{,\phi}$ denotes the deferentiation with respect to $\phi$. 

The field $\phi$ can also be responsible for dark sectors of cosmology. Assuming a massive scalar, $V=m_\phi^2\phi^2/2$ for example, the field $\phi$ behaves as dark matter if $m_\phi\gtrsim10^{-31}~\mathrm{eV}$ and as dark energy for $m_\phi\lesssim10^{-33}~\mathrm{eV}$ \cite{Fujita:2020ecn}. In the dark matter case, the field value changes dynamically before reionization and continues to oscillate with a small amplitude at $z\lesssim10$ \cite{Nakatsuka:2022epj}. In the dark energy case, the field may begin slow-rolling in the relatively late universe. 

The polarization of the CMB is a powerful tool for searching for cosmic birefringence. The total rotation angle measured by the CMB is given by $\beta_0=\beta(z_{\mathrm{LSS}})$, where $z_{\mathrm{LSS}}$ is the redshift at the Last Scattering Surface (LSS). The cross-angular power spectrum between the CMB $E$-modes and $B$-modes is proportional to $\beta_0$ and has been used to measure $\beta_0$ from the CMB data \cite{Zhao:2015mqa, Lue:1998mq,Feng:2006dp}. 

Since we can only measure $\beta(z_{\mathrm{LSS}})$ with CMB, it is challenging to determine the time evolutions $\beta(z)$. Therefore, CMB data alone cannot distinguish between models that predict different time evolutions of $\phi$ at the redshift below $z_{\mathrm{LSS}}$ and are not sensitive to whether $\phi$ behaves as dark energy or dark matter. CMB is only useful for constraining models of cosmic birefringence which $\phi$ evolved during the recombination and reionization epoch \cite{Sherwin2023, Dongdong:2024dmi, Lee:2022udm, Galaverni:2023zhv, Naokawa:2023upt, Naokawa:2024xhn, Murai:2022zur}.  

Another challenge with the CMB is the degeneracy between the birefringence angle $\beta_0$ and the instrumental miscalibraion angle $\alpha$ because what we observe is $\alpha+\beta_0$ \cite{Larson_2011, Planck:2016soo}. Recent studies with Planck and WMAP data have attempted to break this degeneracy by using foreground emission as a calibrator \cite{Minami:2019ruj, Minami:2020xfg}, but concerns about unknown systematics such as mis-modeling of $C_l~{EB}$ from an intrinsic foreground emission still remain. The result of ACT \cite{ACT:2025fju} did not use the foreground calibration technique, and their angle calibration accuracy is insufficient to draw a definitive conclusion. The ongoing ground-based CMB telescope, Simons Observatory \cite{SimonsObservatory:2025wwn}, aims to reduce the calibration uncertainty of $\alpha$ to a sufficiently low level, enabling an absolute measurement of the reported value of $\beta_0$. However, a calibration of $\alpha$ with an uncertainty well below 0.1 deg remains an extremely challenging task. 

Therefore, independent measurements of cosmic birefringence, especially those that do not rely on the CMB, are of great interest \cite{Yin:2024fez}. If a pseudoscalar is the origin of cosmic birefringence, not only CMB photons but also photons of any kind could be affected. The major photon sources, other than the CMB, are astrophysical objects within $z\lesssim10$. To detect the rotation using these sources, we need linearly polarized sources with known intrinsic angles. Hereafter we refer to such sources as ``standard cross". 

Radio galaxies with strong jets originating from the AGN core are candidates for the standard cross. These galaxies typically emit linearly polarized strong synchrotron radiations. When defining the jet position angle as $\chi$ and the polarization position angle as $\psi$, it is known that the relative angle $\theta\equiv\chi-\psi$ is distributed randomly but with a peak at 90~deg \cite{Clarke1980, Joshi2007}. The statistical error of the peak position should be $\hat{\sigma}_\theta\equiv\sigma_\theta/\sqrt{n}$ according to the central limit theorem, where $\sigma_\theta$ is the standard deviation of the distribution of $\theta$, and $n$ is the sample size of radio galaxies. With cosmic birefringence, the peak at an arbitrary redshift should shift to $90+\beta(z)$~deg.  Therefore, we can use radio galaxies as a standard cross with enough samples \cite{Carroll:1989vb, Nodland:1997cc, Carroll:1997tc}.

The very first attempts to detect cosmic birefringence used this correlation in the 1990s \cite{Carroll:1989vb, Nodland:1997cc, Carroll:1997tc}. They also combined the redshift information with $\order{10^2}$ samples of radio galaxies. Although these pioneering works resulted in null detections, as discussed in their paper for future prospects, the CMB has since emerged as a powerful tool for probing cosmic birefringence. Now that CMB provides hints of $\beta_0\sim0.3$ deg and the sample size of radio galaxies has increased \cite{Kuźmicz_2018, Dabhade2020, Simonte:2022rye, Simonte2024, Heinz, Oei2023, Mostert2024}, it is time to reconsider the test using astrophysical sources. This test also plays an important role in the tomographic measurement of $\beta(z)$ at low redshift, enabling not only an independent test of cosmic birefringence but also the constraint of the model parameter of $\phi$. Therefore, we investigate the time evolution of $\beta(z)$ at low-redshift to understand its detectability.

\ifdefined \submitletter
  {\it Dark energy profile.---}
\else
  \section{Dark energy profile} \label{sec:intro}
\fi
The field $\phi$ should remain stationary for $z\gtrsim10$ and begin to evolve in the late universe to detect the rotation of cosmic birefringence from both the CMB and low-redshift sources. This corresponds to the epoch when the pseudoscalar plays the role of dynamical dark energy and is slow-rolling in the simple cases. Under the slow-roll regime, the condition
\begin{equation}
\label{24010721}
    \ddot{\phi} \ll 3H\dot{\phi}.
\end{equation}
is satisfied, meaning that the Hubble friction term is large enough to prevent $\phi$ from accelearating. From Eq.(\ref{24010720}) and Eq.(\ref{24010721}), we derive
\begin{equation}
    3H\dot{\phi} \simeq -V_{,\phi}~~\therefore~~
    d\phi \simeq -\frac{V_{,\phi}}{3H}dt.
\end{equation}
During the slow-rolling, $V_{,\phi}$ is almost constant, and the rotation angle at $z$ is written as
\begin{equation}
    \label{25040401}
    \beta(z)
    \simeq
    \frac{1}{2}g
    V_{,\phi}\int_{t(z=0)}^{t(z)}
    \frac{1}{3H}dt.
\end{equation}
Using $\beta_0=\frac{1}{2}gV_{,\phi}\int^{t(z_{\mathrm{LSS}})}_{t(z=0)}\frac{1}{3H(t)}dt$, the profile is given by
\begin{equation}
\label{25021405}
    \beta(z) = \beta_0 \times 
    \frac{\int_{t(z=0)}^{t(z)}
    \frac{1}{3H}dt}
    {\int^{t(z_{\mathrm{LSS}})}_{t(z=0)}
    \frac{1}{3H}dt}
    ~~\left(\equiv \beta_0\Xi(z)\right).
\end{equation}
Since $g$ and $V(\phi)$ appear both in the numerator and denominator and are taken out of the integral, they canceled out. Therefore, the normalized profile $\Xi(z)$ is explicitly independent of the model parameters of the pseudoscalar field, if it is a spectator field. 

In the flat $\Lambda$CDM model, the time evolution of Hubble parameter is determined as
\begin{equation}
\label{25032901}
    \begin{aligned}
        H_\Lambda(z)
        \simeq
        H_0\sqrt{(1+z)^3(1-\Omega_{\Lambda0})+\Omega_{\Lambda0}}.
    \end{aligned}
\end{equation}
in the flat $\Lambda$CDM. $\Omega_{\Lambda0}$ is the present energy density of the cosmological constant. We neglect the radiation because their contribution is negligible when the pseudoscalar field begins to vary. Using $dz=-(1+z)H(z)dt$ and Eq.(\ref{25032901}), we can derive
\begin{equation}
    \begin{aligned}
        &\int^{t(z)}_{t(z=0)}\frac{1}{H(t)}dt 
        \equiv-\frac{1}{H_0^2}F(z)\\
        &=
        -\frac{1}{H_0^2}\int^z_0\frac{dz}{(1+z)^4(1-\Omega_{\Lambda0})+(1+z)\Omega_{\Lambda0}},
    \end{aligned}
\end{equation}
then the normalized profile is 
\begin{equation}
    \label{25040201}
    \begin{aligned}
        \Xi(z) 
        &= 
        F(z)/F(z_{\mathrm{LSS}})\\
        &= 
        F^{-1}(z_{\mathrm{LSS}})\times
        \int^z_0\frac{dz}{(1+z)^4(1-\Omega_{\Lambda0})+(1+z)\Omega_{\Lambda0}}.
    \end{aligned}
\end{equation}
Substituting $F^{-1}(z_{\mathrm{LSS}})\simeq1.77$ into Eq.(\ref{25021405}), the total birefringence angle of the photons from the source at $z$ is given by
\begin{equation}
    \label{25021701}
    \beta_{\mathrm{DE}}(z) \equiv 1.77\beta_0 
    \int^z_0\frac{dz}{(1+z)^4(1-\Omega_{\Lambda0})+(1+z)\Omega_{\Lambda0}}.
\end{equation}
Eq.(\ref{25021701}) depends only on $\Omega_{\Lambda0}$. Hereafter, we refer to this profile as the ``dark energy profile".

We note that the discussion above assumes that the contribution of $\phi$ to the energy density is negligible. However, when we include this contribution to the energy density, the Hubble parameter becomes dependent on $\phi$, and $\Omega_{\Lambda0}$ is replaced by 
$(1-r) \Omega_{\Lambda0} + r\rho_{c0}^{-1}\left( \dot{\phi}^2/2 + V(\phi) \right)$
, where $r$ is the ratio of the pseudoscalar contribution to the total dark energy density, $\rho_{c0}$ is the critical energy density, and
$\rho_{c0}^{-1}\left( \dot{\phi}^2/2 + V(\phi) \right)|_{z=0} = \Omega_{\Lambda0}$ should be satisfied. When dark energy is fully accounted for by the pesudoscalar ($r=1$), the Hubble parameter is given by
\begin{equation}
    \label{25021406}
    H_\phi(z) = H_0
    \sqrt{(1+z)^3(1-\Omega_{\Lambda0}) + \rho_{c0}^{-1}\left( \frac{1}{2}\dot{\phi}^2 + V(\phi) \right)}.
\end{equation}
For the slow-rolling $\phi$, the time variation of $\phi$ is sufficiently small and $H_\phi(z)\simeq H_{\Lambda}(z)$. Therefore, we expect that the profile of $\beta(z)$ does not depend on the details of the $\phi$ and is uniquely determined by $\Omega_{\Lambda0}$.

\begin{figure}
    \centering
    \includegraphics[width=85mm]{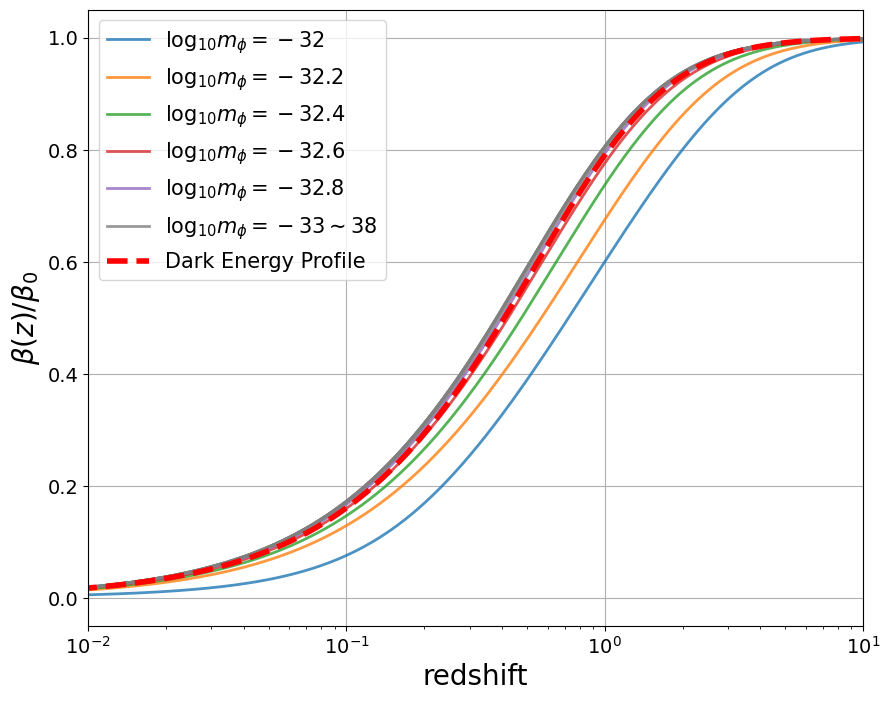}
    \caption{The profiles of cosmic birefringence angle when the potential is mass type, $V(\phi)=m_\phi^2\phi^2/2$ and their comparison to the dark energy profile. The solid lines are the numerically solved profiles with Eq.(\ref{24010720}) and Eq.(\ref{25021406}). The colored ones are when $m_\phi$ is slightly heavier than $10^{-33}$ eV. The grey lines are when $m_\phi=10^{-33},10^{-34},10^{-35},10^{-36},10^{-37},10^{-38}$ eV. The red dotted line is the dark energy profile of Eq.(\ref{25040201}) and is analytically computed.}
    \label{fig:mass}
\end{figure}

To verify these statements, we numerically solve Eq.(\ref{24010720}) and Eq.(\ref{25021406}) with the mass model, $V(\phi)=m_\phi^2\phi^2/2$, shown in Fig.~\ref{fig:mass}. The solid lines represent the numerical results for several mass values smaller than $10^{-32}~\mathrm{eV}$ while the red dotted line shows the dark energy profile obtained from Eq.~(\ref{25021701}). When $m_\phi\gtrsim10^{-33}~\mathrm{eV}$, the profiles deviate from the dark energy profile. On the other hand, for the slow-rolling cases where $m_\phi\lesssim10^{-33}~\mathrm{eV}$, all profiles almost coincide with the dark energy profiles as predicted above. We also check the cases of cosine profile under the similar setup to Ref.\cite{Fujita:2020ecn} and confirmed the similar feature as the result of the case of the mass potential. We further investigate the case of linear potential. We calculated various slope parameters, $s$, defined by Ref.\cite{Gasparotto:2022uqo} and confirm that all cases are well approximated by the dark energy profile\footnote{We note that, in the linear potential case, the dark energy profile is justified without slow-roll condition for the matter dominant era. Eq.(5.4) of Ref.\cite{Gasparotto:2022uqo} without slow-rolling is equivalent to Eq.~(\ref{25040401}). The authors also pointed out that the ratio of $\beta(z_{\mathrm{LSS}})$ and $\beta(z_{\mathrm{reio}})$, where $z_{\mathrm{reio}}$ is the redshift at reionization, does not depend on parameters such as $s$.}.

\ifdefined \submitletter
  {\it Detectability.---}
\else
  \section{Detectability.} \label{sec:intro}
\fi
Previous results provide strong motivation to test cosmic birefringence with astrophysical sources at redshifts of $z\lesssim10$. If we can detect birefringence angles consistent with dark energy profile, it would independently confirm the cosmic birefringence reported from the CMB. Such a detection would simultaneously indicate the detection of dynamical dark energy. If the dark energy profile is rejected with sufficient statistical significance, we can rule out the possibility that cosmic birefringence is induced by slow-rolling dark energy. 

\begin{figure}
    \centering
    \includegraphics[width=85mm]{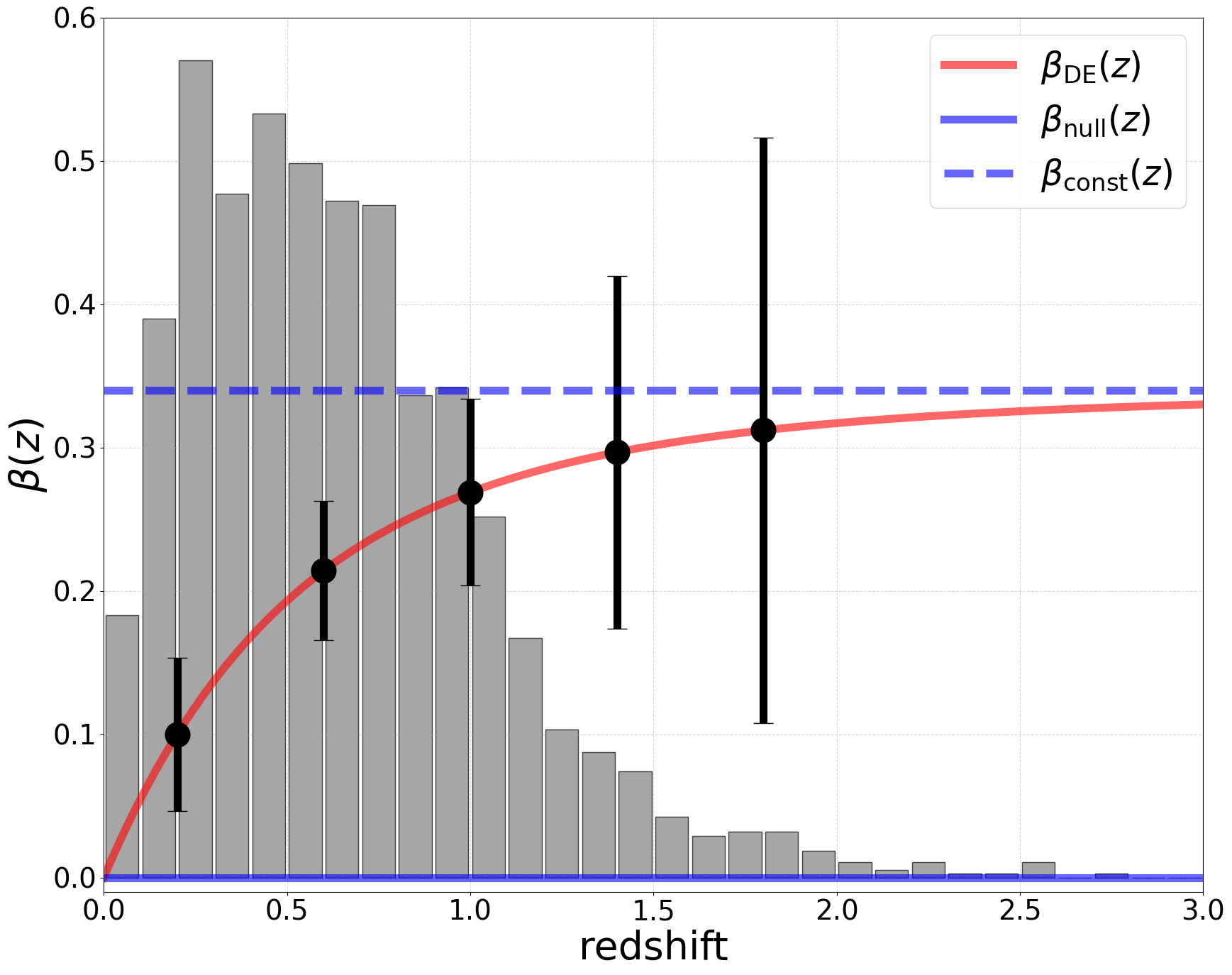}
    \caption{The forecasted error bars when the sample size of radio galaxies is one million and the fiducial profile is the dark energy profile (the red solid line). The blue solid (dotted) line denotes the dark matter (constant) profile. The histgram shows the normalized redshift distribution of polarized sources samples from Ref. \cite{OSullivan:2023eub}.}
    \label{fig:forecast_1}
\end{figure}

We forecast the sample size of radio galaxies needed to perform this test with certain significance levels. We use the following assumptions for simplicity: 
\begin{itemize}[left=0pt]
    \item $\sigma_\theta$ is $\sim30$~deg~\cite{Carroll:1997tc}, and does not depend on the redshift.
    \item The redshift distribution of the samples matches that of LOFAR LoTSS Data Release 2 \cite{OSullivan:2023eub}.
    \item No errors in measuring the source redshift.
    \item The samples are distributed up to $z=2$.
\end{itemize}
Under these assumptions, we evaluated following chi-squared
\begin{equation}
    \label{25022501}
    \chi^2_{\mathrm{DE vs null}} = \sum_i\frac{[\beta_{\mathrm{DE}}(z_i)-\beta_{\mathrm{null}}(z_i)]^2}{\hat{\sigma}_\theta^2},
\end{equation}
\begin{equation}
    \label{25040202}
    \chi^2_{\mathrm{DE vs Const}} = \sum_i\frac{[\beta_{\mathrm{DE}}(z_i)-\beta_{\mathrm{const}}(z_i)]^2}{\hat{\sigma}_\theta^2},
\end{equation}
with $\beta_0=0.34~$deg to estimate the sample size required to distinguish the dark energy profile and the null profile $\beta_{\mathrm{null}}(z)\equiv0$ deg or constant profile $\beta_{\mathrm{const}}(z)\equiv0.34$ deg. We divide the samples into five redshifte bins, where $i$ represents the index of the bins. The redshift of the samples in each bin is represented by the central value of the bin. The results are
\begin{equation}
    \label{25040203}
    \chi^2_{\mathrm{DE vs null}} = 49\left(\frac{\sigma_\theta}{30~\mathrm{deg}}\right)^{-2}\left(\frac{n}{10^6}\right),
\end{equation}
\begin{equation}
    \label{25040204}
    \chi^2_{\mathrm{DE vs Const}} = 28\left(\frac{\sigma_\theta}{30~\mathrm{deg}}\right)^{-2}\left(\frac{n}{10^6}\right).
\end{equation}
We also calculate the $p$-value ($p$) from the result of Eq.(\ref{25040203}) and Eq.(\ref{25040204}) and find that we need about 0.4 (0.7) million samples to distinguish dark energy profile from dark matter profile and 0.7 (1.3) million samples from constant profile with $p=10^{-3(6)}$. Fig.~\ref{fig:forecast_1} shows the dark energy profile with the forecasted error bars.

\ifdefined \submitletter
  {\it Summary \& Discussion.---}
\else
  \section{Summary & Discussion} \label{sec:intro}
\fi
In this Letter, we investigate the low-redshift profile of the cosmic birefringence angle $\beta(z)$ and its application to the independent test of the reported angle $\beta_0\sim0.3~\mathrm{deg}$ from CMB observations. We conclude that the evolution shows a unique profile, introduced as the ``dark energy profile" in Eq.~(\ref{25021701}) when cosmic birefringence is induced by a slow-rolling pseudoscalar field, which is responsible for dynamical dark energy. This profile is determined solely by $\Omega_{\mathrm{\Lambda0}}$ and does not depend on any other parameters or the model of pseudoscalar. We also estimated that $\order{10^5-10^6}$ polarized radio galaxies are sufficient to detect this signal. This method is also useful to distinguish between dark energy and other possible origins of cosmic birefringence including systematics, which is difficult to do with only CMB alone. Therefore, the test of dark energy profile with low-redshift polarized sources is not only an independent method but also a complementary one.  

Recent polarized source surveys by LOFAR LoTSS have identified about 2,500  distant polarized sources from about an eigths sky fraction and revealed that the vast majority, $>90$\% are radio galaxies \cite{OSullivan:2023eub}. Further survey of the entire northern sky by LOFAR will increase the sample size. The ASKAP POSSUM survey \cite{Anderson2021} will increase the sample size by more than 2 orders of magnitude in the near future. SKA \cite{Weltman:2018zrl} is expected to further increase the sample size \cite{George2020}. Other next-generation high-frequency surveys by VLA \cite{Lacy:2019rfe} and DSA-2000 \cite{Hallinan2019DSA} will also improve data quality. Therefore, the sample size suggested above is not unrealistic in the near future. The precise determination of the redshift for each source is also crucial. Recent and upcoming optical galaxy surveys, such as DESI \cite{DESI2018}, PFS \cite{PFSTeam:2012fqu}, LSST \cite{LSST}, Euclid \cite{Euclid} will play a key role in achieving this.

Recent work on cosmic birefringence signal from CMB, Ref~\cite{Naokawa:2024xhn}, pointed out that we cannot exclude the possibility of $\beta_0=0.34+180n$ deg with non-zero integer $n$. Tomographic investigation of low-redshift profiles with radio galaxies are useful to break the degeneracy. Also, $\beta(z)$ becomes much larger in the case of nonzero $n$, then we can exclude or detect with much smaller samples. We note that a larger $n$ will require a more precise redshift determination because the sign of $\beta(z)$ flips more times between -90 and 90 degrees. The cases of $n\sim \order{1}$ can be tested with existing data.

Although we used the correlation of jet angle and polarization angle as the standard cross for the forecast, there are other candidates for standard crosses. The correlation between the intensity gradient and the polarization direction for well-resolved lobes of radio galaxies was proposed as a standard cross in the 1990s \cite{Leahy:1997wj}. Another example is a recently proposed method that utilizes the correlation between the semimajor axis and the integrated polarization angle of spiral galaxies. Regardless of the types of standard crosses, we can combine all samples for the profile test \cite{Yin:2024fez}.

There may be concerns that some astrophysical effects or the morphologies of each sample could introduce systematic errors. However, these uncertainties are already taken into account in the variance $\sigma^2_\theta$ of the distribution of $\theta$. We expect that no astrophysical effect will induce global parity violation within the framework of standard physics. Therefore, the systematic errors can be eliminated with sufficient sample statistics. Another concern may be that the instrumental miscalibration angle $\alpha$ could cause serious systematics. If all samples are simultaneously biased by $\alpha$, we cannot eliminate it statistically. However, when testing the dark energy profile, we are primarily focusing on the curvature as a function of redshift. Therefore, we can separate the two effects as $\alpha + \beta_{\mathrm{DE}}(z)$ as we have already shown by distinguishing $\beta_{\mathrm{DE}}(z)$ and $\beta_{\mathrm{Const}}(z)$.

Lastly, we comment on the case of dynamical dark energy, explaining recent DESI results \cite{DESI:2025zgx}. Several studies have already shown that the (pseudo)scalor field, including the cases of $m_\phi\sim\order{10^{-33}}$ eV \cite{Luu:2025fgw, Urena-Lopez:2025rad}, can explain the result and that is compatible with cosmic birefringence \cite{Yin:2024hba, Tada2024, Lee:2025yvn, Nakagawa:2025ejs}. Therefore, the dark energy profile is expected to be useful for examining the DESI results. Even if the pseudoscalar field is not ``slow"-rolling, and $\beta(z)$ can deviate from the dark energy profile, $\beta(z)$ is expected to be nonzero at $z\lesssim10$, which means that the test using astrophysical sources can still be effective to detect or exclude dynamical dark energy. Furthermore, tests with low-redshift sources are also useful for searching other possible origins such as domain walls \cite{Hiramatsu:2012sc, Ferreira:2023jbu}, which can produce cosmic birefringence at low-redshift. Investigating specific profiles for them is an important direction for future work. 


\begin{acknowledgments}
We thank Jun'ichi Yokoyama for his guidance of the research activitiy regarding this Letter. We are grateful to Kohei Kamada, Toshiya Namikawa, Ippei Obata, Kai Murai, Maresuke Shiraishi, Ricardo Ferreira, Silvia Gasparotto and Eiichiro Komatsu for fruitful discussion on this paper. We also would like to thank Shane O'Sullivan, Takuya Akahori, Yoshiyuki Inoue, Kotaro Kohno, Rikuto Omae, Shinsuke Ideguchi, Kohei Kurahara, Haruka Sakemi, Martijn Oei and other ASKAP \& LOFAR members for their useful comments and supports especially in respect of radio surveys. ChatGPT was very helpful to complete this work, especially for its coding and language support. We thank the Forefront Physics and Mathematics Program to Drive Transformation (FoPM), a World-leading Innovative Graduate Study (WINGS) Program, the University of Tokyo. We are also supported by JSPS KAKENHI Grant Number 24KJ0668.

\end{acknowledgments}

\appendix


\bibliographystyle{apsrev}
\bibliography{cite}

\end{document}